\documentstyle[12pt]{article}
\def\be{\begin{equation}}
\def\ee{\end{equation}}
\def\bea{\begin{eqnarray}}
\def\eea{\end{eqnarray}}

\begin{document}

\begin{center}
{\Large{\bf Resistance of a Rotating-Moving Brane
with Background Fields Against Collapse}}

\vskip .5cm
{\large Davoud Kamani}
\vskip .1cm
{\it Physics Department, Amirkabir University of Technology
(Tehran Polytechnic)}\\
P.O.Box: {\it 15875-4413, Tehran, Iran}\\
e-mail: {\it kamani@aut.ac.ir}\\
\end{center}

\begin{abstract}

Using the boundary state formalism we investigate the 
effect of tachyon condensation process on a rotating 
and moving D$p$-brane with various background 
fields in the bosonic string theory. The rotation and motion 
are inside the brane volume. 
We demonstrate that some specific rotations and/or 
motions can preserve the brane from instability and  
collapse.

\end{abstract}



\vskip .5cm

\newpage

\section{Introduction}

Some significant steps have been made to
study the D-branes as essential objects
in the string theory and some of their specifications such 
as stability \cite{1,2}. The instability 
of the branes can be studied by 
the open string tachyon dynamics and tachyon condensation 
phenomenon \cite{3}. The unstable D-branes decay into 
the closed string vacuum or to some lower dimensional unstable 
branes as intermediate states \cite{4,5,6}. These intermediate 
states also decay to lower dimensional stable configurations or 
to the closed string vacuum. 
These concepts have been studied by various methods, e.g. 
the string field theory \cite{4,7,8}.

On the other hand, there is the boundary state method for 
describing D-branes \cite{9} - \cite{16}. This method is 
an applicable tool in many complicated situations.
Thus, this valuable formalism can be applied for investigating 
the tachyon condensation process 
\cite{17} - \cite{20}. For example, 
the boundary state is a source for closed strings, 
therefore, by using this state and the tachyon condensation, 
one can find the time evolution of the source.
In addition, it has been argued that the boundary state description 
of the rolling tachyon is valid during the finite 
time which is determined by the 
string coupling, and the energy of the system could be dissipated
to the bulk beyond this time \cite{19}.
Besides, this method elucidates the 
decoupling of the open string modes
at the non-perturbative minima of the tachyon potential \cite{21}.

In this article we consider a rotating and 
moving D$p$-brane in the presence
of a $U(1)$ gauge potential in the worldvolume of the brane and a 
tachyon field. The rotation and linear motion of the brane 
will be considered in its volume.
Presence of the above background fields 
indicates some preferred alignments within the brane. 
Thus, we shall demonstrate that the Lorentz 
symmetry on the worldvolume of the brane has been broken, 
and hence such rotations and motions
are sensible. The boundary state corresponding to
this non-stationary D$p$-brane 
enables us to investigate the tachyon condensation for 
gaining a new understanding of this phenomenon on the D-branes. 
In fact, we shall observe that condensation of tachyon 
cannot always impel the brane to be unstable. 
In other words, by considering a rotating-moving brane, the 
reduction of the brane dimension sometimes does not occur, and hence 
in spite of the tachyon condensation process we have a stable brane.

This paper is organized as follows. In Sec. 2, the boundary
state corresponding to a dynamical D$p$-brane 
with various background fields 
(in the context of the bosonic string theory) will be constructed. 
In Sec. 3, stability of this D-brane 
under the condensation of the tachyon will be investigated. 
Section 4 is devoted to the conclusions.
%%%%%%%%%%%%%%%%%%%%%%%%%%%%%%%%%%%%%%%%%%%%%%%%%%%%%%%%%%
\section{The boundary state of the brane}

As we know open strings live on the D-branes. This elucidates that 
the corresponding fields of the open string states, such as the 
gauge potential $A_\alpha (X)$ and the tachyon field $T(X)$, 
exist on the worldvolume of a D$p$-brane. Since the boundary of the 
emitted (absorbed) closed string worldsheet sits on the brane, 
the closed string possesses some interactions with 
the open string fields. In other words, the open string fields 
naturally behave as backgrounds for any closed string which is emitted 
(absorbed) by the brane. However, in various papers 
the open string fields $A_\alpha (X)$ and $T(X)$ have been 
extremely applied as valuable backgrounds for studying 
the closed strings, e.g. see Refs. 
\cite{4, 8, 16, 17, 18, 19, 20, 21, 23, 24}.

For constructing the boundary state associated with a 
non-stationary D$p$-brane in the presence of the above background 
fields, we start with the action
\bea
S = &-& \frac{1}{4\pi\alpha'} {\int}_\Sigma
d^{2}\sigma \sqrt{-h}h^{ab}g_{\mu\nu}\partial_a X^{\mu}\partial_b
X^{\nu}
\nonumber\\
&+& \frac{1}{2\pi\alpha'} {\int}_{\partial\Sigma}
d\sigma \left( A_\alpha
\partial_{\sigma}X^{\alpha}+ 2\omega_{\alpha\beta}
X^\alpha \partial_\tau X^\beta
+T(X^{\alpha}) \right),
\eea
where $\Sigma$ is a closed string worldsheet, emitted 
(absorbed) by
the brane, and $\partial\Sigma$  is its boundary.
The set $\{X^\alpha|\alpha =0, 1, \cdot \cdot \cdot ,p \}$ 
represents the worldvolume directions and the set
$\{X^i| i= p+1, \cdot \cdot \cdot ,d-1\}$ indicates 
the directions perpendicular to it.
This action includes 
the $U(1)$ gauge potential $A_\alpha$ which lives in the 
brane worldvolume, a tachyonic field $T(X)$ and 
a dynamical term. This term has the spacetime 
angular velocity ${\omega }_{\alpha \beta}$ for the brane rotation 
and motion in its volume. The components  
$\{\omega_{0 {\bar \alpha}}|{{\bar \alpha} =1, \cdot\cdot\cdot,p}\}$ 
specify the speed of the brane and the elements 
$\{\omega_{{\bar \alpha}{\bar \beta}}
|{\bar \alpha},{\bar \beta} =1, \cdot\cdot\cdot,p\}$ show 
its rotation. 

Unlike the closed string tachyon and dilaton backgrounds which 
appear in the bulk of the string action and, because of their 
couplings with the two-dimensional curvature, break the Weyl 
symmetry our tachyon belongs to the open string spectrum, 
hence it specifies a surface term for the string action. Therefore, 
this tachyon field does not couple to the two-dimensional curvature.  
This fact enables us to choose the flat gauge for the worldsheet 
metric, e.g. see Ref. \cite{4, 8, 19, 24, 25}. However, beside 
this choice, for simplification of the calculations the spacetime metric
is chosen as $g_{\mu \nu}=\eta_{\mu \nu}=
{\rm diag} (-1,1,\cdot\cdot\cdot,1)$, and for
the gauge potential we use the gauge
$A_{\alpha}=-\frac{1}{2}F_{\alpha \beta }X^{\beta}$
where the field strength $F_{\alpha \beta }$
is constant. Furthermore, we apply the following tachyon
profile $T=T_0 +\frac{1}{2}U_{\alpha\beta}X^{\alpha}X^{\beta}$
where $T_0$ and the symmetric matrix $U_{\alpha\beta}$ are constant.

Note that the second term of the action in the elected gauge
can be written as 
$\frac{1}{4}F_{\alpha \beta} J^{\alpha \beta}_\sigma$.
Thus, $F_{\alpha \beta}$ comes from the brane,
i.e. electric and magnetic fields on the brane, while 
$J^{\alpha \beta}_\sigma$ belongs to the closed string.
The third term of the action, which possesses the feature 
$\omega_{\alpha \beta}J^{\alpha \beta}_\tau$, is an analog of the  
second one, i.e. ${\omega }_{\alpha \beta}$ is an effect of 
the brane while $J^{\alpha \beta }_{\tau}$ pertains
to the closed string.

The action gives the following boundary state equations
\bea
&~& \left( M_{\alpha \beta } {\partial }_{\tau }X^{\beta }
+F_{\alpha \beta}{\partial
}_{\sigma }X^{\beta } +U_{\alpha \beta }X^{\beta }\right)_{\tau =0}
|B\rangle=0 ,
\nonumber\\
&~& ({\delta X}^i)_{\tau =0}|B\rangle=0 ,
\eea
where 
$M_{\alpha \beta}=\eta_{\alpha \beta} + 4\omega_{\alpha \beta}$.

Now we prove that the background fields break the Lorentz 
invariance along the worldvolume of the brane. 
The action of the Lorentz generators on the boundary state 
can be extracted from Eq. (2),
\bea
J^{\alpha \beta}|B\rangle
= \int^\pi_0 d \sigma \bigg{[}
(M^{-1}F)^\alpha_{\;\;\;\gamma} X^\beta
\partial_\sigma X^\gamma
- (M^{-1}F)^\beta_{\;\;\;\gamma} X^\alpha
\partial_\sigma X^\gamma
\nonumber\\
+ (M^{-1}U)^\alpha_{\;\;\;\gamma} X^\beta X^\gamma
- (M^{-1}U)^\beta_{\;\;\;\gamma} X^\alpha X^\gamma
\bigg{]}_{\tau =0}|B\rangle.
\eea
This equation elaborates that for restoring the Lorentz
symmetry the tachyon matrix $U_{\alpha \beta}$
and the field strength $F_{\alpha \beta}$ 
should vanish. We observe that even in the absence of the 
electromagnetic fields, the tachyon field independently  
breaks the Lorentz invariance along the brane worldvolume.
However, since the right-hand-side of Eq. (3) is a functional of the 
spacetime coordinates along the brane worldvolume we deduce 
that the Lorentz symmetry breaking is local. That is, the linear 
motion of the brane in any direction obviously is sensible. 
The same also is true for its rotation. 
For obtaining more perception of rotation and motion
of a D$p$-brane in its volume, beside the Lorentz invariance
breaking, we should reminisce that such configurations accurately 
are T-dual of some imaginable systems. More precisely, a
rotating-moving D$p$-brane can also be constructed via T-duality 
from a D$(p-1)$-brane which rotates and moves perpendicular 
and parallel to its volume. For example, consider a D1-brane 
along the $x^1$-direction with the velocity
components $V^1$ and $V^2$ in the directions $x^1$ and $x^2$,
respectively. Now apply the T-duality in
the $x^2$-direction that we assume it is compact. 
The resulted system represents a D2-brane 
along the $x^1x^2$-plane with the velocity $V^1$ in the
$x^1$-direction and possesses an electric field $E=V^2$ 
in the $x^2$-direction. Similarly, it is possible to 
recast a rotating D2-brane via the T-duality from another
setup of the D1-brane. Note that after accomplishing 
the T-duality effects one can decompactify all compact 
coordinates.
                                  
The other interpretation of Eq. (3) is
as follows. The tensor operator $J^{\alpha \beta}$ 
defines angular momentum of the closed string  
state $|B\rangle$ along the brane worldvolume. 
According to the action (1) since the closed string couples to 
the background fields and spacetime angular velocity of the 
brane, its angular momentum also is affected by these variables,
as expected. 

In terms of the closed string oscillators Eqs. (2) find the feature 
\bea
&~& \left[\left(M_{\alpha \beta}
- F_{{\mathbf \alpha }{\mathbf \beta }} +\frac{i}{2m}U_{\alpha
\beta }\right){\alpha }^{\beta }_m +{\left(M_{\alpha \beta}
+F_{{\mathbf \alpha } {\mathbf
\beta }}\ -\frac{i}{2m}U_{\alpha \beta }\right)}{\widetilde{\alpha
}}^{\beta }_{-m}\right] {|B\rangle}^{\left({\rm osc}\right)}\ =0 ,
\nonumber\\
&~& \left( 2\alpha'M_{\alpha \beta} p^{\beta }
+U_{\alpha \beta }x^{\beta
}\right) {|B\rangle}^{\left(0\right)}\ =0,
\nonumber\\
&~& ({\alpha }^{i}_m-{\widetilde{\alpha }}^{i}_{-m})
{|B\rangle}^{\left({\rm osc}\right)}\ =0,
\nonumber\\
&~& (x^{i}-y^{i}){|B\rangle}^{\left(0\right)}\ =0,
\eea
where the transverse vector 
$\{y^i| i= p+1, \cdot \cdot \cdot, d-1\}$ 
defines the location of the brane. The boundary state is 
given by the direct product 
$|B \rangle={|B\rangle}^{\left(0\right)}
\otimes {|B\rangle}^{\left({\rm osc}\right)}$.

The solution of the zero mode part of the boundary state is given by 
\bea
{{\rm |}B\rangle}^{\left(0\right)}
&=& \frac{1}{\sqrt{\det (U/2)}}\int^{\infty }_{{\rm -}\infty }
\exp\bigg{\{}i{\alpha }^{{\rm '}}\bigg{[}\sum^{p}_{\alpha  =0}
{\left(U^{{\rm -}{\rm 1}}M\right)}_{\alpha \alpha}
{\left(p^{\alpha}\right)}^{{\rm 2}}
\nonumber\\
&+& \sum^{p}_{\alpha ,\beta {\rm =0},\alpha \ne \beta}{{\left(U^{{\rm -}{\rm
1}}M+M^T U^{-1}\right)}_{\alpha \beta }
p^{\alpha }p^{\beta}}\bigg{]}\bigg{\}}
\left( \prod_{\alpha}{\rm |}p^{\alpha}\rangle dp^{\alpha}\right)
\otimes\prod_i{\delta {\rm (}x^i}{\rm -}y^i{\rm )}
{\rm |}p^i{\rm =0}\rangle .
\eea  
The factor $1/\sqrt{\det (U/2)}$ is induced by the 
disk partition function \cite{22}.
The exponential factor of this state, which is absent in the 
conventional boundary states, originates from the tachyon field.
For the oscillating part
the coherent state method imposes the following solution 
\bea
{|B\rangle}^{\left({\rm osc}\right)}\
=\frac{T_p}{g_s}\prod^{\infty }_{n=1} {[\det
Q_{(n)}]^{-1}}\;{\exp \left[-\sum^{\infty }_{m=1}
{\frac{1}{m}{\alpha }^{\mu }_{-m}S_{(m)\mu \nu } {\widetilde{\alpha
}}^{\nu }_{-m}}\right]\ } {|0\rangle}\;,
\eea
where the matrices have the following definitions 
\bea
&~& Q_{(m){\alpha \beta }} = M_{\alpha \beta}
-F_{{\mathbf \alpha }{\mathbf \beta
}}+\frac{i}{2m}U_{\alpha \beta },
\nonumber\\
&~& S_{(m)\mu\nu}=\left( \Delta_{(m)\alpha \beta}
\; ,\; -{\delta}_{ij}\right) ,
\nonumber\\
&~& \Delta_{(m)\alpha \beta} = (Q_{(m)}^{-1}N_{(m)})_{\alpha \beta},
\nonumber\\
&~& N_{(m){\alpha \beta }} = M_{\alpha \beta}
+F_{{\mathbf \alpha }{\mathbf \beta }}
-\frac{i}{2m}U_{\alpha \beta }.
\eea
The mode dependency of these matrices is induced by the 
tachyon matrix. The normalization factor
$\prod^{\infty }_{n=1}{{[\det Q_{(n){\alpha \beta }}]}^{-1}}$
is inspired by the disk partition function \cite{22}.
The zeta function regulation gives
$\sum_{n=1}^\infty 1 \rightarrow -1/2$, and hence  
$\prod_{n=1}^\infty \det U^{-1} \rightarrow \sqrt{\det U}$.
Therefore, for the total prefactor of the boundary state we acquire 
\bea
\frac{\prod^{\infty}_{n=1} [\det Q_{(n)}]^{-1}}
{\sqrt{\det (U/2)}} \longrightarrow  2^{(p+1)/2}
\prod^{\infty}_{n=1}\det \left[ (M -F)U^{-1} 
+\frac{i}{2n} {\mathbf 1}
\right]^{-1}\;.
\eea

According to the first equation of (4),
on the boundary state one can express
the right-moving oscillator $\alpha^\alpha_m$ in terms of 
all left-moving oscillators 
$\{{\tilde \alpha}^\beta_{-m}|\beta = 0,1, \cdot\cdot\cdot,p \}$, 
which causes the matrix $\Delta_{(m)\alpha \beta}$ to appear in 
the boundary state. If we express ${\tilde \alpha}^\alpha_m$ 
in terms of the set 
$\{\alpha^\beta_{-m}|\beta = 0,1, \cdot\cdot\cdot,p \}$, 
the boundary state possesses the matrix 
$\left( [\Delta^{-1}_{(-m)}]^T\right)_{\;\alpha \beta}$. 
Equality of these matrices gives the condition  
$\Delta_{(m)}\Delta^T_{(-m)}={\mathbf 1}$. Since this 
equation comprises all mode numbers of the closed string
it splits to the following conditions
\bea
&~& \eta U -U\eta +4(\omega U+ U\omega ) =0 ,
\nonumber\\
&~& \eta F - F \eta 
+4(\omega F +F\omega)=0,
\eea
which are independent of the mode numbers. These equations
reduce $(p+1)(3p+2)/2$ parameters of the theory to
$p(p+1)/2$.  

In fact, the total boundary state contains a portion 
of the conformal ghosts too. Since this part does not include  
the background fields it will not contribute to the 
tachyon condensation. Thus, in the future discussions
we shall neglect it.
%%%%%%%%%%%%%%%%%%%%%%%%%%%%%%%%%%%%%%%%%%%%%%%%%%%%%%%%%%%%%%%
\section{Tachyon condensation and brane stability}

Study of the open string tachyon field which 
began basically by the Sen's papers \cite{3} demonstrates that this 
tachyon field has an essential role on improving our
knowledge about the fate of the D-branes, their instability 
or stability, true vacuum of the tachyonic string theories 
and so on. In fact, the tachyonic modes
of the string spectrum make the D-branes to be unstable \cite{3}.
As the tachyon condenses the dimension of the
brane decreases and in the final stage only the closed string
degrees of freedom remain.
From the boundary sigma-model point of view, in the 
$d$-dimensional spacetime the tachyon condensation 
starts by a conformal theory with the $d$ Neumann 
boundary conditions in the UV fixed point. Then adding the tachyon 
as a perturbation (deformation) will cause the theory 
to roll toward an IR fixed point. Afterwards we receive a closed 
string vacuum with a D$p$-brane, which corresponds to a new 
vacuum with $(d-p-1)$ Dirichlet boundary conditions.

Since the boundary state represents a brane in
terms of the quantum states of closed string and 
contains the disk partition function as the normalization factor,
it is an appropriate tool for evaluating 
the behavior of a D$p$-brane 
under the tachyon condensation process. Therefore,
by applying this formalism, 
in this section we shall demonstrate that for a
dynamical brane, some rotations and/or motions can
prevent the brane from instability and collapse.
This is different from the conventional tachyon
condensation which reduces the brane dimension.

For making the tachyon condensation, 
some of the tachyon's components (at least one of them) 
should go to infinity. In Eqs. (5) and (6) there are the matrices  
$L=U^{-1}M+M^T U^{-1}$, 
$\;(U^{-1}M)_{\alpha \alpha}$ which is diagonal,
$\Delta_{(m)}$, and also the prefactor 
in the right-hand side of Eq. (8)  
in which the tachyon matrix $U_{\alpha \beta}$ exists. 
That is, we should take the limits of these 
variables to obtain the evolution of the D$p$-brane under the 
condensation of the tachyon.
Now let the system approach to the infrared fixed point, i.e.   
$U_{pp} \to \infty$. This gives the following limit
\bea
{\mathop{\lim }_{U_{pp}\to \infty }
(U^{-1})_{p\alpha }=\mathop{{\rm \lim}}_{U_{pp}\to \infty }
(U^{-1})_{\alpha p}=0\ }, \;\;\;\alpha = 0,1, \cdot\cdot\cdot, p .
\eea

The Eq. (10) induces the following limit on the 
elements of the diagonal matrix 
\bea 
(U^{-1}M)_{\alpha \alpha}
=(U^{-1})_{\alpha \gamma'}M^{\gamma'}\; _{\alpha}
+(U^{-1})_{\alpha p}M^p \;_{\alpha}
\longrightarrow
(U^{-1})_{\alpha \gamma'}M^{\gamma'} \;_{\alpha}\;,
\;\;\gamma' \in \{0, 1,\cdot\cdot\cdot ,p-1 \}.
\eea
Thus, the element $(U^{-1}M)_{pp}$ vanishes, and 
the diagonal matrix $(U^{-1}M)_{\alpha \alpha}$
reduces to a $p\times p$ diagonal matrix
$(U^{-1})_{\alpha' \gamma'}M^{\gamma'}\;_{\alpha'}$
with $\alpha' , \gamma' \neq p$. In fact, this diagonal matrix 
is not a main portion of the  boundary state. Hence 
reduction of it to a $p\times p$ matrix is ignorable. 
We shall see that all other matrices, in the infrared fixed point, 
remain $(p+1)\times (p+1)$.

The matrix elements 
$L_{\alpha \beta}|_{\alpha \neq \beta}$ have the limit
\bea
L_{\alpha \beta}=(U^{-1})_{\alpha \gamma'}M^{\gamma'}\;_\beta
+(M^T)_\alpha \;^{\gamma'}\;(U^{-1})_{\gamma' \beta}
\;\;\;\;,\;\;\;\alpha \neq \beta .
\eea
Therefore, in this limit the elements $L_{\alpha' p}=L_{p \alpha'}
=4 (U^{-1})_{\alpha'}\; ^{\gamma'}\;
\omega_{\gamma' p} $ are nonzero unless the case
$\{\omega_{\gamma' p}=0 \;|\gamma'=0,1,\cdot\cdot\cdot,p-1\}$. 
Since $\omega_{0 p}$ is corresponding
to the velocity of the brane along the $x^p$-direction and
$\omega_{\gamma' p}|_{\gamma' \neq 0 ,p}$ associated with the
brane rotation inside the $x^{\gamma'}x^p$-plane we observe
that these components of the motion and rotation preserve this
portion of the boundary state 
against dimensional reduction. This also elaborates
that the other components of the rotation and motion, i.e.
$\{\omega_{\alpha' \beta'}| \alpha' , \beta' =0,1,\cdot\cdot\cdot,p-1\}$,
do not preserve this portion of the boundary state 
against collapse. We shall observe
that the dimensional preserving also occurs for the
other parts of the boundary state via the same components 
of the rotation and motion.

The prefactor of the boundary state, i.e. 
the right-hand side of Eq. (8), under the limit $U_{pp} \to \infty$, 
takes the feature  
\bea
2^{(p+1)/2}
\prod^{\infty}_{n=1}\det \left[ (M -F){\tilde U}^{-1} 
+\frac{i}{2n} {\mathbf 1}\right]^{-1}\;.
\eea
According to Eq. (10) the matrix ${\tilde U}^{-1}$ 
is similar to $U^{-1}$ where its 
last row and its last column possess zero matrix elements.
This implies that the last column of the matrix 
$(M -F){\tilde U}^{-1}$
vanishes while its last row has the nonzero elements 
$(4\omega -F)_p \;^{\gamma'}({\tilde U}^{-1})_{\gamma' \beta}$.
That is, after the process of the tachyon condensation the 
normalization factor of the boundary state also respects 
the totality of the D$p$-brane.

Now the matrix $\Delta_{(m)\alpha \beta}$ is investigated. 
In the limit $U_{pp} \to \infty$
the last row of this matrix vanishes 
except the element $\Delta_{(m)pp}$ which tends to
$-1$. The elements of the last column, which are 
$(\Delta_{(m)})_{\alpha' p}|_{\alpha' \neq p}$, remain nonzero if 
$\{\omega_{\alpha' p} \neq 0 |\alpha'=0,1,\cdot\cdot\cdot,p-1\}$.
Thus, this part of the boundary state also resists
against dimensional reduction of the brane.

Adding all these together 
we observe that the rotation of the brane in the planes 
$\{x^{\bar \alpha}x^p|{\bar \alpha} = 1,2, \cdot\cdot\cdot,p-1\}$ 
and/or its motion along 
the direction $x^p$ induce a resistance against the instability 
and collapse of the brane. However, the other components of 
the brane dynamics do not preserve it.

{\bf Example: the D2-brane}

For illustrating the stability of a non-stationary  
brane under the tachyon condensation phenomenon consider 
the simplest example, i.e. the D2-brane. The tachyon condensation 
is applied by the limit $U_{22} \to \infty$. Therefore, 
according to Eq. (13) the dimension of the matrices in the 
total prefactor remains $3 \times 3$. Besides, 
the matrix elements $L_{02}$ and $L_{12}$ don't vanish.
In addition, we obtain
\bea
{\mathop{\lim }_{U_{pp}\to \infty }}
\Delta_{(m)}=\left( \begin{array}{ccc}
\Delta_{(m)00} &
\Delta_{(m)01} &
\Delta_{(m)02}\\
\Delta_{(m)10} &
\Delta_{(m)11} &
\Delta_{(m)12}\\
0 & 0  & -1
\end{array} \right),
\eea
where
\bea
&~& \Delta_{(m)02}
=\frac{-8 \omega_{12}(4\omega_{01}-F_{01}+\frac{iU_{01}}{2m})
+8 \omega_{02}(1+\frac{iU_{11}}{2m})}
{(-1+\frac{iU_{00}}{2m})(1+\frac{iU_{11}}{2m})
+(4 \omega_{01}-F_{01})^2+(\frac{U_{01}}{2m})^2}\;,
\nonumber\\
&~& \Delta_{(m)12}
=\frac{-8 \omega_{02}(-4\omega_{01}+F_{01}+\frac{iU_{01}}{2m})
+8 \omega_{12}(-1+\frac{iU_{00}}{2m})}
{(-1+\frac{iU_{00}}{2m})(1+\frac{iU_{11}}{2m})
+(4 \omega_{01}-F_{01})^2+(\frac{U_{01}}{2m})^2}\;.
\nonumber
\eea
All the above facts elucidate that motion of the brane along the 
$x^2$-direction and/or its rotation in the $x^1x^2$-plane 
preserve it against instability. We observe that motion
of the brane along the $x^1$-direction does not have this ability.
%%%%%%%%%%%%%%%%%%%%%%%%%%%%%%%%%%%%%%%%%%%%%%%%%%%%%%%%%%
\section{Conclusions and outlook}

We studied effects of tachyon 
condensation on a rotating-moving D$p$-brane with the 
$U(1)$ gauge potential 
and the quadratic tachyon field via the boundary state 
method in the bosonic string theory. The conventional tachyon 
condensation usually is terminated by reduction of the brane 
dimension. In this article we observed that special rotations 
and/or special motions of the brane can protect it from 
instability and hence dimensional reduction. As it was proved, 
because of some components of the linear velocity and angular
velocity of the brane, at the infrared fixed point it resists 
against the instability and collapse. More precisely, this dynamics 
of the brane imposes a solitonic behavior to it. 
However, the other rotations and motions of the brane in its 
volume do not preserve it.

It is valuable and interesting to extend the setup of this 
article to the transverse linear motion and a rotation in which 
one of the brane directions to be axis of rotation.
Besides, the extension can be done for the BPS and 
non-BPS D-branes in the superstring theory for the 
tangential rotation and motion, and also for the transverse
rotation and motion through the boundary state formalism.
%%%%%%%%%%%%%%%%%%%%%%%%%%%%%%%%%%%%%%%%%%%%%%%%%%%%%%%%%%%%%%


\begin{thebibliography}{99}

\bibitem{1}
J. Polchinski, ``{\it String Theory}'', (Cambridge University Press,
Cambridge, 1998), Volumes I and II; 
C.V. Johnson, ``{\it D-Branes}'', (Cambridge
University Press, Cambridge, 2003).
\bibitem{2}
J. Polchinski, Phys. Rev. Lett. {\bf 75} (1995) 4724.
\bibitem{3}
A. Sen, Int. J. Mod. Phys. {\bf A 14} (1999) 4061; 
Int. J. Mod. Phys. {\bf A 20} (2005) 5513; 
JHEP {\bf 9808} (1998) 010; 
JHEP {\bf 9808} (1998) 012;
JHEP {\bf 9812} (1998) 021; 
JHEP {\bf 9809} (1998) 023;
JHEP {\bf 9910} (1999) 008; 
JHEP {\bf 9912} (1999) 027;
M. Frau, L. Gallot, A. Lerda and P. Strigazzi, Nucl. Phys. {\bf B 564} 
(2000) 60.
\bibitem{4}
D. Kutasov, M. Marino and G. Moore, JHEP {\bf 0010} (2000) 045.
\bibitem{5}
K. Hashimoto, P.M. Ho and J.E. Wang, Mod. Phys. Lett. 
{\bf A 20} (2005) 79.
\bibitem{6}
T. Lee, Phys. Rev. {\bf  D 64} (2001) 106004; 
Phys. Lett. {\bf B 520} (2001) 385; T. Lee, K.S. 
Viswanathan and Y. Yang, J. Korean Phys. Soc. {\bf 42} (2003) 34.
\bibitem{7}
P. Kraus and F. Larsen, Phys. Rev. {\bf D 63} (2001) 106004.
\bibitem{8}
E. Witten, Phys. Rev. {\bf D 47} (1993) 3405;
Phys. Rev. {\bf D 46} (1992) 5467.
\bibitem{9}
M.B. Green and P. Wai, Nucl. Phys. {\bf B 431} (1994) 131;
M. Li, Nucl. Phys. {\bf B 460} (1996) 351;
C. Schmidhuber, Nucl. Phys. {\bf B 467} (1996) 146.
\bibitem{10}
M.B. Green and M. Gutperle, Nucl. Phys. {\bf B 476} (1996) 484.
\bibitem{11}
M. Billo, P. Di Vecchia and D. Cangemi, Phys. Lett. {\bf B 400} (1997) 63.
\bibitem{12}
F. Hussain, R. Iengo and C. Nunez, Nucl. Phys. {\bf B 497} (1997) 205.
\bibitem{13}
O. Bergman, M. Gaberdiel and G. Lifschytz, Nucl. Phys. {\bf B 509} (1998) 194.
\bibitem{14}
P. Di Vecchia, M. Frau, I. Pesando, S. Sciuto, A. Lerda and R. Russo, Nucl.
Phys. {\bf B 507} (1997) 259.
\bibitem{15}
M. Billo, P. Di Vecchia, M. Frau, A. Lerda, I. Pesando, R. Russo
and S. Sciuto, Nucl. Phys. {\bf B 526} (1998) 199.
\bibitem{16}
H. Arfaei and D. Kamani, Phys. Lett. {\bf B 452} (1999) 54,
arXiv: hep-th/9909167;
Nucl. Phys. {\bf B 561} (1999) 57, arXiv: hep-th/9911146;
Phys. Lett. {\bf B 475} (2000) 39, arXiv: hep-th/9909079; 
D. Kamani, Mod. Phys. Lett. {\bf A 15} (2000) 1655, arXiv: hep-th/9910043;
Phys. Lett. {\bf B 475} (2000) 39, arXiv: hep-th/9909079;
Nucl. Phys. {\bf B 601} (2001) 149, arXiv: hep-th/0104089;
Phys. Lett. {\bf B 487} (2000) 187, arXiv: hep-th/0010019;
F. Safarzadeh-Maleki and D. Kamani, Phys. Rev. {\bf D 89}, 026006 
(2014), arXiv: 1312.5489 [hep-th]; 
Phys. Rev. {\bf D 90}, 107902 (2014), arXiv: 1410.4948 [hep-th];
J. Exp. Theor. Phys. {\bf 119} (2014) 677, arXiv: 1406.2667 [hep-th];  
Z. Rezaei and D. Kamani, J. Exp. Theor. Phys. {\bf 113} (2011) 956,
arXiv: 1106.2097 [hep-th];
J. Exp. Theor. Phys. {\bf 114} (2012) 234,
arXiv: 1107.1183 [hep-th].

\bibitem{17}
A. Sen, JHEP {\bf 0204} (2002) 048; JHEP {\bf 0207} (2002) 065.
\bibitem{18}
F. Larsen, A. Naqvi and S. Terashima, JHEP {\bf 0302} (2003) 039. 
\bibitem{19}
T. Okuda and S. Sugimoto, Nucl. Phys. {\bf B 647} (2002) 101.
\bibitem{20}
M. Naka, T. Takayanagi and T. Uesugi, JHEP {\bf 0006} (2000) 007.
\bibitem{21}
G. Chalmers, JHEP {\bf 0106} (2001) 012.
\bibitem{22}
E.S. Fradkin and A.A. Tseytlin, Phys. Lett. {\bf B 163} (1985) 123.
\bibitem{23}
J. Dai, R.G. Leigh and J. Polchinski, Mod. Phys. Lett. {\bf A4} 
(1989) 2073; 
R.G. Leigh, Mod. Phys. Lett. {\bf A 4} (1989) 2767;
S.J. Rey and S. Sugimoto, Phys. Rev. {\bf D 68} (2003) 026003;
S.P. de Alwis, Phys. Lett. {\bf B 505} (2001) 215;
M. Frau, I. Pesando, S. Sciuto, A. Lerda and R. Russo, Phys.
Lett. {\bf B 400} (1997) 52;
C.G. Callan and I.R. Klebanov, Nucl. Phys. {\bf B 465} (1996) 473;
M. Li, Nucl. Phys. {\bf B 460} (1996) 351.
\bibitem{24}
G. Arutyunov, A. Pankiewicz and B. Stefanski Jr, JHEP 0106
(2001) 049.
\bibitem{25}
S.J. Rey and S. Sugimoto, Phys. Rev. {\bf D 67} (2003) 086008;
E.T. Akhmedov, M. Laidlaw and G.W. Semenoff, Pis'ma
Zh. Eksp. Teor. Fiz. {\bf 77} (2003) 3, [JETP Lett. {\bf 77} (2003) 1];
JHEP {\bf 0311} (2003) 021; M. Laidlaw, arXiv: hep-th/0210270; 
M. Laidlaw, Ph.D. thesis, University of British Columbia, 
2003; arXiv:hep-th/0309055.
 
\end{thebibliography}
\end{document}